# Quantum Annealing for Enhanced Feature Selection in Single-Cell RNA Sequencing Data Analysis

Selim Romero[1,2,3], Shreyan Gupta[1,3], Victoria Gatlin[1,3], Robert S. Chapkin[2,3], and James J. Cai[1,3,4*]

[1]Department of Veterinary Integrative Biosciences, Texas A&M University, College Station, TX 77843, USA.
[2]Department of Nutrition, Texas A&M University, College Station, TX 77843, USA.
[3]CPRIT Single Cell Data Science Core, Texas A&M University, College Station, TX 77843, USA.
[4]Department of Electrical and Computer Engineering, Texas A&M University, College Station, TX 77843, USA.

Corresponding Author: James J. Cai

Keywords: Quantum annealing, quadratic unconstrained binary optimization (QUBO), feature selection, scRNA-seq, quantum computing

## Abstract
Feature selection is a machine learning technique for identifying relevant variables in classification and regression models. In single-cell RNA sequencing (scRNA-seq) data analysis, feature selection is used to identify relevant genes that are crucial for understanding cellular processes. Traditional feature selection methods often struggle with the complexity of scRNA-seq data and suffer from interpretation difficulties. Quantum annealing presents a promising alternative approach. In this study, we implement quantum annealing-empowered quadratic unconstrained binary optimization (QUBO) for feature selection in scRNA-seq data. Using data from a human cell differentiation system and an anticancer drug resistance study, we demonstrate that QUBO feature selection effectively identifies genes whose expression patterns reflect critical cell state transitions associated with differentiation and drug resistance development. Our findings indicate that quantum annealing-powered QUBO reveals complex gene expression patterns potentially missed by traditional methods, thereby enhancing scRNA-seq data analysis and interpretation.

## Introduction
Single-cell RNA sequencing (scRNA-seq) has transformed our understanding of cellular heterogeneity by providing a detailed view of gene expression at the individual cell level. This technology has enabled unprecedented exploration of gene expression programs that govern cell fate and regulate various cellular processes. However, dissecting molecular mechanisms underlying cellular processes remains a daunting task due to the complexity of gene function. A single gene may be involved in multiple cellular processes, and different genes often interact within intricate regulatory networks. Functionally similar genes may compensate for one another, leading to genetic redundancy, which further complicates the identification of key genes involved in specific cellular processes.

Feature selection is a machine learning technique used to identify a subset of input variables that are most relevant to a target variable. In single-cell research, feature selection is critical for identifying informative genes that capture essential biological insights while reducing data complexity, thereby enhancing the interpretability of biological questions. This study addresses the feature selection problem in scRNA-seq data for regression. Our objective is to identify a subset of genes that combinedly predicts cell state, in which gene expression serves as numerical input, and cell state as a continuous numerical target. Our focus is on regression, albeit feature selection is also employed for classification tasks, such as selecting genes for cell type identification (Yang *et al*. 2021). The least absolute shrinkage and selection (LASSO) is a popular method for feature selection (Tibshirani 1996). LASSO can be used within embedded methods in other applications to reduce the complexity of the single-cell data, making it an efficient, interpretable, and effective method for handling high-dimensional data (Yang *et al*. 2021). However, LASSO is limited to linear models and may not capture nonlinear relationships. Thus, there is a clear need for new effective solutions given that conventional optimization methods e.g., LASSO and other linear regression methods, may struggle with the high dimensionality and nonlinearity inherent in scRNA-seq data, and may become trapped in local minima, potentially missing critical features. Random forest regression (RFR) is a widely used machine learning technique known for its ability to model complex, nonlinear relationships. However, it suffers from several limitations, particularly in interpretability and computational efficiency. As an ensemble of decision trees, it functions as a black-box model, making it difficult to extract meaningful insights from individual predictions. Its high computational cost, especially with large datasets and deep trees, can be a bottleneck in real-time applications.

Quantum annealing has emerged as a viable tool to tackle complex problems in data analysis (Arai *et al*. 2023). This work explores the feasibility of using currently available quantum computer architectures to achieve quantum feature selection in single-cell data analysis. Specifically, we consider feature selection through a quadratic unconstrained binary optimization (QUBO) model, designed to identify features that are both independent and influential. Quadratic optimization is known to scale exponentially with the number of features, which typically poses a significant computational challenge. However, implementing QUBO on quantum annealers can offer a substantial speedup and an increased likelihood of finding the global minimum by exploring the solution space simultaneously through an adiabatic process. By harnessing the power of quantum annealing, QUBO-based feature selection may provide a more effective solution to the regression problems in scRNA-seq data analysis. To this end, we adapt the method proposed by (Mücke *et al*. 2023) and develop a quantum feature selection framework tailored to scRNA-seq data. This framework aims to address the limitations of traditional methods by capturing complex, nonlinear relationships in gene expression that are crucial for understanding cellular processes.

# Results

## Framework of QUBO feature selection for regression

We tackled a regression task using a scRNA-seq dataset $\boldsymbol{X}$ consisting of $p$ cells and $n$ genes, with a target variable $\boldsymbol{T}$ representing the cell state to be predicted. The feature selection problem was framed as identifying a subset of these $n$ genes that can achieve performance comparable to the original dataset for regression tasks. This was accomplished by solving an optimization problem depicted in **Fig. 1**, where the objective was to find the optimal feature set (vector $\boldsymbol{F}^*$) that minimized the QUBO cost function (matrix $\boldsymbol{Q}(\boldsymbol{F}, \alpha; \boldsymbol{I}, \boldsymbol{R})$), accounting for both the importance and redundancy of features. The solution $\boldsymbol{F}^*$ is a binary vector representing the selected features (genes), where $F_i^* = 1$ indicates that the $i$-th feature is selected. The implementation leverages the parameter $\alpha$ to balance the importance and redundancy in constructing QUBO matrix $\boldsymbol{Q}$ for annealing on a quantum computer. This approach ensures that the most informative genes are captured, while minimizing redundancy and enhancing the interpretation and efficiency of the resulting gene set for regression analysis. To estimate feature importance and redundancy for constructing a single cost function, we followed a previous study (MÜCKE *et al.* 2023) and adopted the mutual information. This framework can be adapted by altering the interaction matrix using different importance and measurements such as information entropy and Pearson's correlation, or by incorporating prior knowledge into the matrix $\boldsymbol{Q}$.

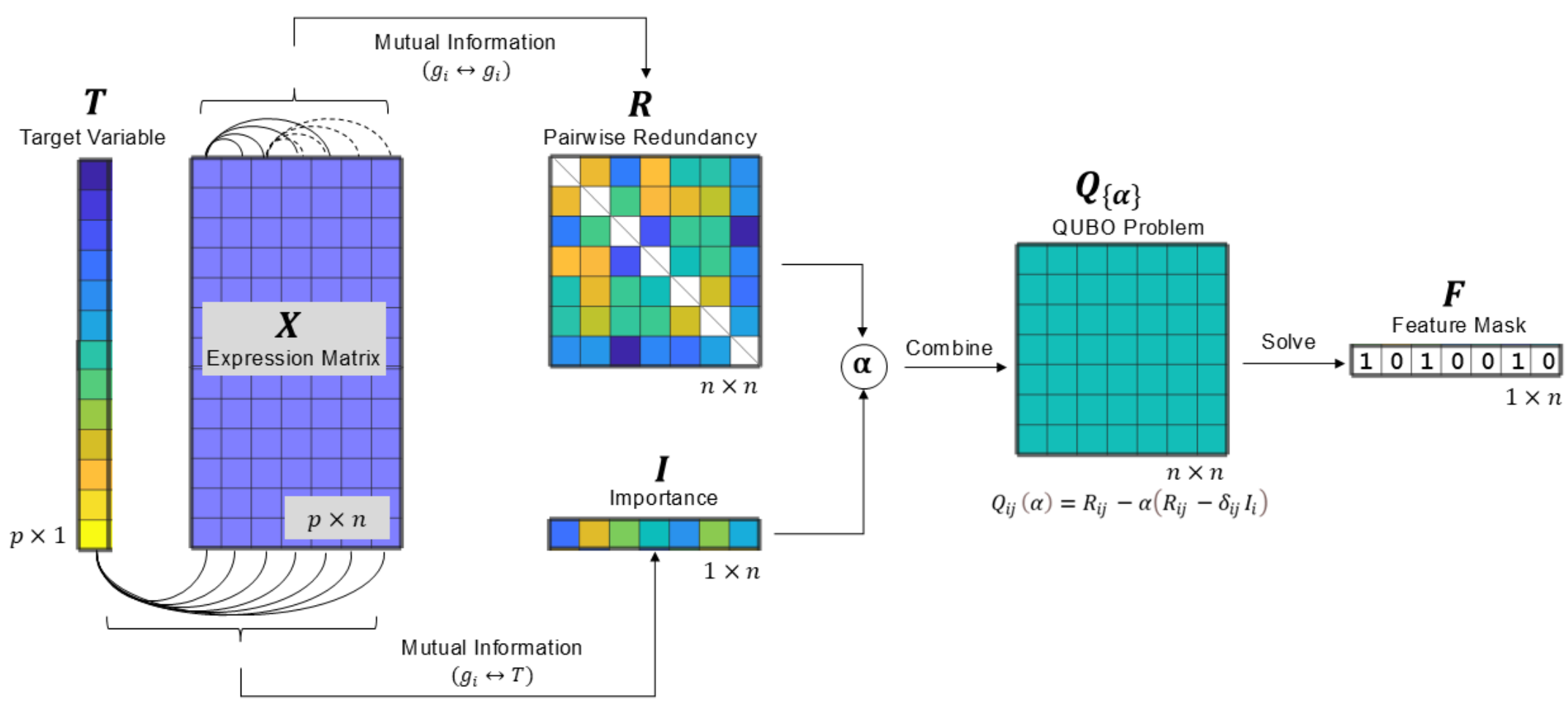

**Fig. 1. Quantum annealing-based QUBO feature selection for regression in scRNA-seq data analysis.** The process begins with an scRNA-seq gene expression matrix $\boldsymbol{X} \in \mathbb{R}^{p \times n}$, where $p$ represents the number of cells and $n$ the number of genes, along with a provided cell state vector $\boldsymbol{T}$ (the target variable). Mutual information is used to compute the redundancy matrix $\boldsymbol{R}$ and the importance vector $\boldsymbol{I}$. A balancing parameter $\alpha$ is then applied to weight importance against redundancy, resulting in the QUBO matrix $\boldsymbol{Q}$. The QUBO is subsequently solved using a quantum annealer, producing a binary solution vector $\boldsymbol{F}^*(\boldsymbol{x}^*)$, which acts as a bitmask to indicate the selected features.

## Solving QUBOs using quantum annealing

Quantum annealing leverages quantum mechanics, specifically utilizing superposition and tunneling effects, to solve optimization problems. In this study, we employed the quantum annealer from D-Wave Systems, specifically designed to address QUBO problems, accessed via the Ocean software development kit, a Python library that interfaces with the quantum annealer. The annealer uses qubits, which can exist in a superposition state of 0 and 1, enabling the simultaneous exploration of many potential solutions. These qubits are superconducting loops, controlled by electric currents and magnetic fields, which create an "embedding landscape" on the chip to guide the optimization process. To validate the results of quantum annealing, we compared the features selected by the D-Wave quantum annealer through the Ocean software development kit with those selected using simulated annealing. The latter employed an iterated tabu search algorithm (PALUBECKIS 2006), implemented in the MATLAB quantum computing package. In all tested cases, whether using quantum or simulated annealing, the results were consistent. Overall, the results obtained from the quantum annealer were validated by the independent classical QUBO solver, confirming that the quantum annealer's selections were fully corroborated by classical methods.

**Table 1. Performance and computational time of different feature selection methods.** Comparative analysis of feature selection methods, showcasing their accuracy and computational time on a synthetic dataset with 10,000 observations and 50 features. The computations were conducted on an OptiPlex SFF plus 7010 PC with 13th Generation Intel® Core™ i5-13500 (24 MB cache, 14 cores, 20 threads, 2.50 GHz to 4.80 GHz turbo, 65 W) and 32 GB RAM.

| Method | Accuracy (%) | Computational Time (s) |
|---|---|---|
| LASSO | 20 | 0.53 |
| Elastic net | 20 | 0.09 |
| RReliefF | 60 | 10.23 |
| Random forest regression (RFR) | 80 | 10.04 |
| Minimum redundancy maximum relevance | 0 | 0.03 |
| Sequential forward feature selection | 80 | 156.15 |
| QUBO (this study) | 100 | 9.80 |

## Simulation analysis of performance of QUBO feature selection vs. other methods

To numerically validate the effectiveness of quantum annealing for feature selection, we conducted a simulation study using synthetic data. We began by generating a normally distributed random matrix $B$ of size $n \times n$ with $n = 50$ features. From this, we computed the correlation matrix $R$ derived from the covariance matrix $C = B^T B$. We then generated multivariate normal data $X$ consisting of $p = 10{,}000$ observations and $n$ features, with a mean $\mu = 0$ and a covariance matrix $\sigma = R$. Correlations were introduced between source features with indices $s = [5, 11, 7, 1, 14]$ and corresponding target features with indices $t = [16, 17, 18, 19, 20]$ using the relation $X_{i,t} = X_{i,s} + \rho\epsilon_i$, where $\rho = 0.1$ is a correlation coefficient, and $\epsilon_i$ is a random normal noise term for the $i$-th observation. The target variable $y$ was constructed using a nonlinear function that depends on the source features:

$$y_i(X_{i,s}) = 0.5\cos(7X_{i,s_4}) + \sin(X_{i,s_3}X_{i,s_2}) + 0.1\exp(X_{i,s_5})\log_2|10X_{i,s_1}| + \rho_1\epsilon_i',$$

where $\epsilon_i'$ is an additional random noise term. The function $y_i$ was designed to be highly nonlinear, with synthetic data influencing the target function. We normalized the target variable $\hat{y}$ to the range $[0, 1]$ and applied z-score standardization to the data $\hat{X}$. We applied QUBO feature selection to the standardized data $\hat{X}$ and predictor $\hat{y}$ to identify features, aiming to recover the original source features $s$. The QUBO model successfully identified the features $[14, 1, 11, 7, 5]$, recovering all source features with a $100\%$ success rate in this simulated scenario. This result indicates the QUBO model's robustness in detecting nonlinear relationships embedded in the predictors. We also attribute this performance to the combined influence of mutual information and balanced redundancy-importance setting, enhancing predictive power in complex problems.

With the same synthetic dataset, we compared the QUBO results with those obtained using six other methods, namely LASSO, elastic net, random forest regression (RFR), RReliefF, sequential forward feature selection, and minimum redundancy maximum relevance (**Table 1**). This diverse array of methods, encompassing linear models, ensemble technique, distance-based heuristic process, and iterative approach, provides a robust and multifaceted benchmark for assessing the performance of QUBO feature selection. We compared the accuracy of selected features and the computational time of different algorithms. As shown in **Table 1**, the QUBO feature selection remained the only one achieving 100% accuracy, followed by RFR and sequential forward feature selection. The high computational time of sequential forward feature selection makes it prohibitive in real application. LASSO and elastic net performed similarly. The LASSO model identified the features $[14, 1, 2, 3, 4]$, achieving only a 40% success rate. Thus, in this simulated scenario, the QUBO feature selection method outperformed all other methods by more accurately identifying relevant features, suggesting that QUBO feature selection may be particularly effective in identifying functionally relevant genes in data derived from complex biological systems.

Based on these results, we selected LASSO and RFR as the primary benchmarks against QUBO in subsequent real-data analyses.

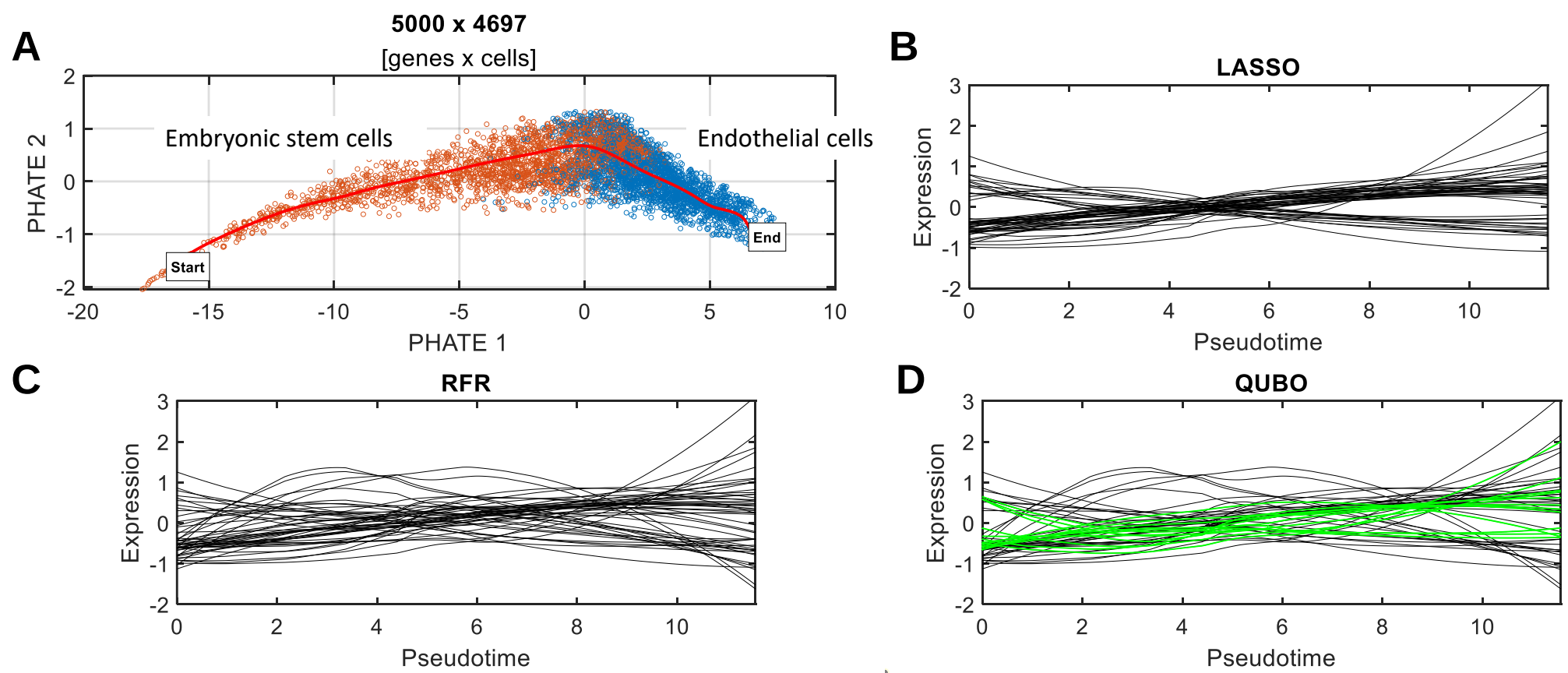

**Fig. 2. Comparative feature selection methods in stem cell-to-endothelial cell differentiation. A.** PHATE projection of scRNA-seq data, with cells color-coded by inferred cell type. The red curve represents the pseudotime trajectory tracking the transition from human embryonic stem cells (hESCs) to endothelial cells (ECs). **B.** Locally weighted linear regression (LOESS)-smoothed expression profiles of the top 50 genes identified by LASSO regression. **C.** Expression profiles of top genes identified through RFR. **D.** Expression profiles of top genes identified by QUBO feature selection. Green lines indicate genes uniquely detected by the QUBO method that were overlooked by LASSO and RFR.

## QUBO feature selection identifies key genes implicated in cell differentiation

We applied QUBO feature selection to a published scRNA-seq dataset (Xu *et al.* 2023), comprising human embryonic stem cells (hESCs) and endothelial cells (ECs). The ECs were derived from the hESCs using the FLI1-PKC system—a high-efficiency induction method for studying cell differentiation (ZHAO *et al.* 2018). After preprocessing the obtained scRNA-seq data (see **Methods** for details), we used PHATE—a visualization method that captures both local and global nonlinear structure in scRNA-seq data (MOON *et al.* 2019)—to embed cells into a 2D latent space (**Fig. 2A**). Subsequently, pseudotime trajectory inference was performed using the splinefit method (CAI 2019), to construct a curve representing the path of cell transition from hESCs to ECs. Here, single-cell data is considered as a snapshot of a continuous process, and the trajectory is reconstructed by finding paths through cellular space that minimize transcriptional changes between neighboring cells (LUECKEN AND THEIS 2019). The ordering of cells along these paths is described by a pseudotime variable. Each cell was projected onto the trajectory and assigned with a pseudotime value. These estimated pseudotime values were utilized as the target variable $\boldsymbol{T}$, for which regression models were constructed.

We applied LASSO, RFR, and QUBO feature selection methods to the data and selected the top 50 genes each for their respective regression model. The expression profiles of these genes were plotted against the pseudotime of hESC-EC differentiation (**Fig. 2B,C,D**).

To understand the function of QUBO-selected genes as a whole set, we conducted gene function enrichment analysis using Enrichr (KULESHOV *et al.* 2016). The top 50 genes identified by QUBO were found to be significantly involved in the following biological processes: *regulation of smooth muscle cell proliferation* (GO:0048660), *regulation of vascular associated smooth muscle cell proliferation* (GO:1904705), and *blood vessel morphogenesis* (GO:0048514), as well as in pathways: *tight junctions* (KEGG term for protein complexes forming semi-permeable connections between ECs), *EPH-ephrin signaling pathway* (R-HSA-2682334), *VEGF-mediated signaling*, and *signaling by GPCR* (R-HSA-372790) (**Supplementary Table 1**).

The top 50 genes selected by LASSO exhibited patterns of monotonic increases or decreases over time in the profile figure (**Fig. 2B**). Out of the 50, 22 genes: DYNLT1, ID1, APLN, THY1, CLDN7, EIF2S2, EVA1B, TMA7, KRT10, PDAP1, FABP5, GNG5, RPS19BP1, ANP32E, ARPC3, HMGB3, PRR13, LITAF, BEX1, DPYSL2, SFRP1 and GDF15, were overlapped with the QUBO top 50 genes. Many of these overlapping genes are known to be implicated in the cellular process of hESC-EC differentiation. For example, APLN, derived from Apelin, regulates the EC development and promotes vascular repair (MASOUD *et al.* 2020). ID1 directly upregulates VEGF, promoting the proliferation and EC migration (QIU *et al.* 2022). THY1 (CD90), commonly known as a stemness marker due to its role in cellular growth and development, is actively involved in angiogenesis (LEE *et al.* 1998).

The top 50 genes selected by RFR exhibited more nonlinear expression profiles over time (**Fig. 2C**). Out of the 50, 34 genes were overlapped with the QUBO top 50 genes. These genes are: DYNLT1, WHAB, ID1, APLN, THY1, CLDN7, EIF2S2, EVA1B, TMA7, KRT10, TRH, PDAP1, FABP5, VIM, ATF5, RPS19BP1, ANP32E, KLK10, ARPC3, MAP1B, POMP, TUBA1C, IGFBP5, S100A3, HMGB3, SLC4A11, PCAT14, BEX1, SFRP1, SPCS3, TP53I11, NTS, GDF15 and CD81. Many of them are known to be functionally relevant cell differentiation. For example, IGFBP5 has been implicated in stem cell differentiation and is known to play a role in angiogenesis and vascular development. It regulates vascular EC functions by modulating the availability of insulin-like growth factors, which are essential for EC proliferation, migration, and survival (Song *et al.* 2024). MAP1B, primarily known for its role in neuronal development, also plays a role in EC function and angiogenesis (Harding *et al.* 2017). SFRP1 plays a crucial role in Wnt signaling, which is vital for EC proliferation and vascular development (Olsen *et al.* 2017). VIM is an EC marker, playing an essential role in maintaining the structural integrity of blood vessels (Boraas and Ahsan 2016), and also facilitates EC migration (Ridge *et al.* 2022).

Among the top 50 genes identified by QUBO, 12 are unique, not overlapping with the top 50 genes from LASSO or RFR. These exclusively QUBO-identified genes are: ID3, ACTR2, MYL12B, IER2, DRAP1, XRCC5, EIF4EBP1, NAP1L1, RAMP2, RGS10, HMGN1, and OAZ1 (highlighted with green curves in **Fig. 2D**). Two genes of particular interest are ID3 and RAMP2, which are critically involved in EC processes. ID3, a member of the inhibitor of DNA binding protein family, encodes a protein that inhibits differentiation and regulates cell cycle progression and cellular differentiation, including angiogenesis (Pammer *et al.* 2004; Osinski *et al.* 2022). RAMP2, a key component of the CRLR/RAMP complex, mediates adrenomedullin's angiogenic effects on ECs (Kamitani *et al.* 1999; Fernandez-Sauze *et al.* 2004). Notably, most of these genes are ranked lower than 100 in the LASSO-identified gene list, and their rankings in the RFR-identified genes vary significantly, particularly as a function of input parameter K—the number of required features.

The results indicate that QUBO feature selection reveals genes with significant overlap with those identified by LASSO and RFR. The expression profiles of these selected genes demonstrate both linear and nonlinear correlations with pseudotime, i.e., the target variable. Furthermore, QUBO uniquely identifies genes not detected by LASSO and RFR. These novel gene selections are functionally pertinent to cell differentiation, highlighting the distinctive effectiveness of QUBO feature selection in this biological analysis.

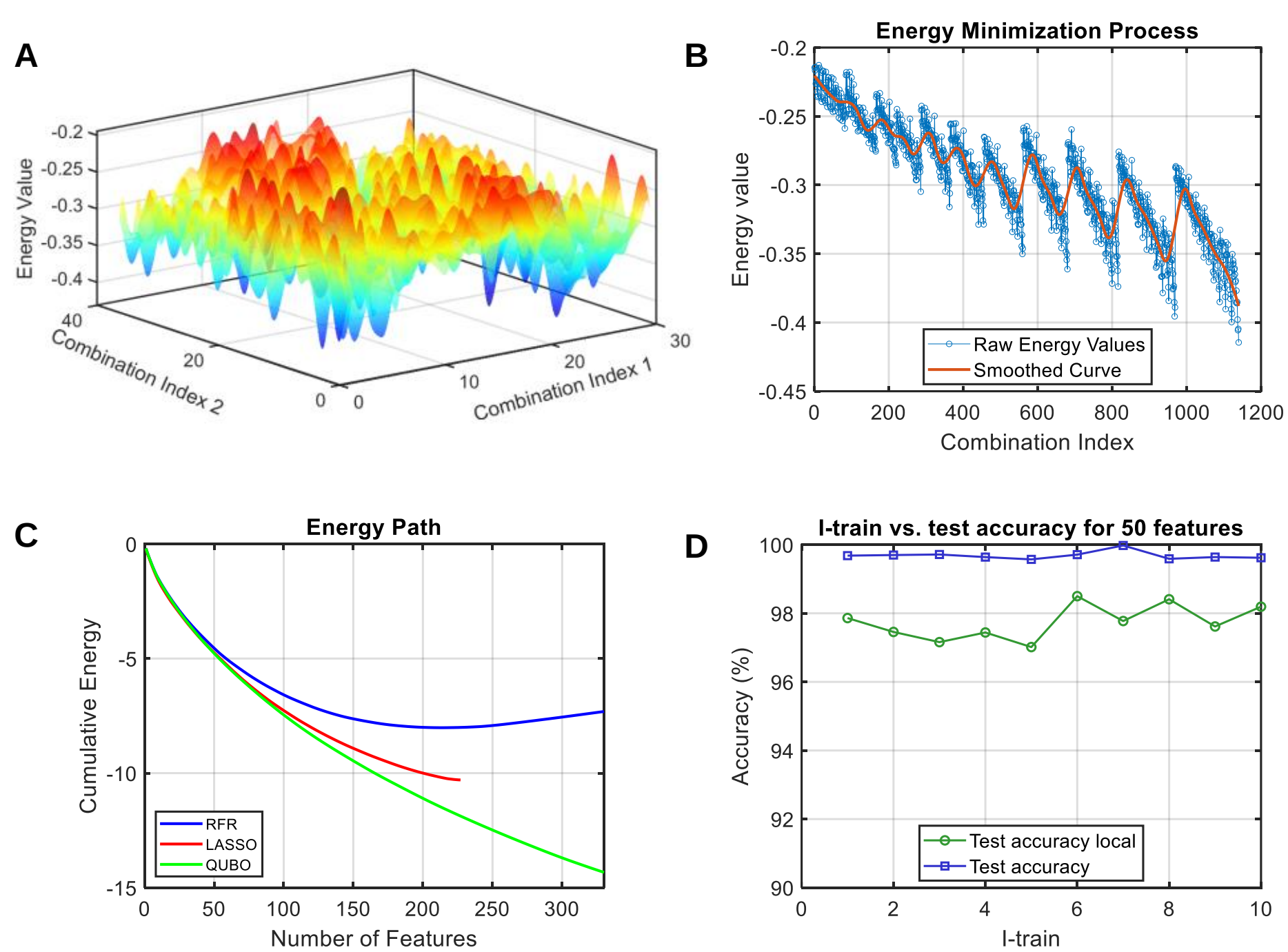

**Fig. 3. Energy landscape of the multiconfigurational feature selection search space.** The landscape depicts energy values for different combinations of features selected from the top 20 genes, which were identified from an initial pool of the top 50 genes based on importance

and redundancy. **A.** 3D visualization of the energy landscape for combinations of 3 features. Feature combination indices are mapped to a 2D grid, with the corresponding energy values on the Z-axis. **B.** Raw energy values for all possible combinations of 3 features. The combinations were sorted according to their energy contribution to the final solution, revealing a more structured pattern compared to if they were plotted in no particular order. A smoothed curve highlights the complex fluctuations of the cost function. **C.** Energy path comparison of QUBO against RFR and LASSO feature selection for the top 500 features. The energy path represents the cumulative energy contribution of sequentially selected features, akin to climbing a ladder where each step represents a feature's energy score. This visualization illustrates the ability of the QUBO cost function to effectively minimize energy, leading to the selection of unique feature combinations and avoiding co-regulated features, thereby facilitating the identification of master regulators. LASSO suggests that after approximately 200 features, further features are not needed to explain the dataset, thus ceasing feature selection. On the other hand, QUBO identifies features that provide an energetically favorable state unreachable by either LASSO or RFR. **D.** Robustness QUBO feature selection through 10-Fold cross-validation with local and global cost function. This plot illustrates the accuracy of the selected features across 10 folds using both, a local and a global cost function. Notably, accuracy achieved by the global cost function $Q$ and the local cost function $Q'$ consistently exceeds 97%.

## Solution quality, stability, and robustness of QUBO feature selection

Quantum annealing offers a powerful approach for feature selection by exploring the energy landscape of a defined Hamiltonian. This Hamiltonian represents the total energy associated with different feature combinations, enabling the exploration of a multiconfigurational space. As the number of features increases, this space expands exponentially, making traditional methods computationally challenging. **Fig. 3** illustrates the complexity of this landscape and the efficacy of our QUBO-based approach. Specifically, **Fig. 3A** represents a 3D visualization of the energy landscape for all possible combinations ($n$ = 1,140) of three out of the top 20 QUBO-selected feature genes in the hESC-EC differentiation dataset, highlighting the complex fluctuations of the cost function and the presence of numerous local minima. **Fig. 3B** displays an unfolded 2D version of the 1,140 combinations and corresponding raw energy values. This 2D plot, where combinations are sorted by their energy contribution, reveals a more structured pattern that underscores the robustness and solution quality achieved by the QUBO framework. These visualizations highlight the challenges faced by traditional optimization techniques in navigating such complex energy landscapes, where the rugged topology of the cost function often traps these methods. In contrast, our QUBO solver effectively balances redundancy and importance, leveraging the quantum annealing process to effectively explore the landscape and identify high-quality solutions, as demonstrated by the structured pattern in **Fig. 3B**.

To further evaluate our approach, we compared the energy paths of solutions obtained from QUBO, LASSO and RFR, all evaluated within our QUBO cost function (**Fig. 3C**). This analysis, performed by calculating the energy path $E(x') = (x')^T Q x'$ (see Methods section for details), where $x'$ represents the binarized feature selection vector from each method, and $Q$ is the QUBO matrix, highlighting the advantage of incorporating both feature importance and redundancy into our cost function, enabling the identification of key "master regulator" features in single-cell data. In the context of solution configurations, each feature's energy score defines its contribution to the total QUBO cost function. The sequential feature selection of features helps visualize the energy landscape in **Fig. 3C**, although the annealing process involves a multiconfigurational evaluation rather than a strictly sequential one. **Fig. 3C** captures this process, where the energy path represents the cumulative effect of selecting features. LASSO struggles to assign meaningful coefficients to many features beyond a certain threshold, which is reflected in **Fig. 3C** where its energy path plateaus. LASSO's limitation becomes more pronounced with larger feature sets, where its energy path flattens abruptly, this behavior was observed as well with the simulated data, where LASSO's performance declined with higher complexity in the dataset's predictor. RFR showed good performance with small feature sets but is prone to selecting redundant features in larger sets (**Supplementary Fig. S1**). In contrast, QUBO reliably scales to larger feature sets, as demonstrated by its consistently decreasing energy path in **Fig. 3C**. Importantly, QUBO prioritizes features with the greatest impact on minimizing the cost function, ensuring that the most important features are consistently selected first. This stability is evident in the consistent ranking of top features, such as the top 20 remaining consistent even when selecting a larger feature set (e.g., 100 features), providing a robust and reliable method for feature ranking and selection.

To assess the robustness of our QUBO feature selection framework, we employed a 10-fold cross-validation and stability analysis. Our approach leverages the direct embedding of the solution vector into the cost function, enabling the evaluation of alternative solutions within the problem's matrix. Critically, the full problem's QUBO matrix, denoted as $Q$, is constructed using the entire dataset. For each fold, we created a subset of the dataset from the count matrix ($X' \in X$) and recomputed a local cost function, denoted as $Q'$, specifically from this subset. We then solved the subset $Q'$ and evaluated the solution $x'$ obtained from the fold against both the local cost function $Q'$ and the global cost function $Q$ (**Fig. 3D**).

Our results confirm the robustness of QUBO feature selection, even in the presence of single-cell data heterogeneity. As shown in **Fig. 3D**, the local cost function $Q'$ effectively maintains a balance between redundancy and importance within each fold. Notably, the global cost function $Q$ achieves an accuracy level exceeding 97% when evaluated against the solutions from each fold, highlighting the consistency and reliability of QUBO-derived solutions across different data partitions. Thus, the QUBO framework effectively navigates complex energy landscapes, delivers high-quality solutions, and demonstrates robustness under cross-validation. It outperforms traditional machine learning methods, particularly in high-dimensional feature selection for single-cell data analysis.

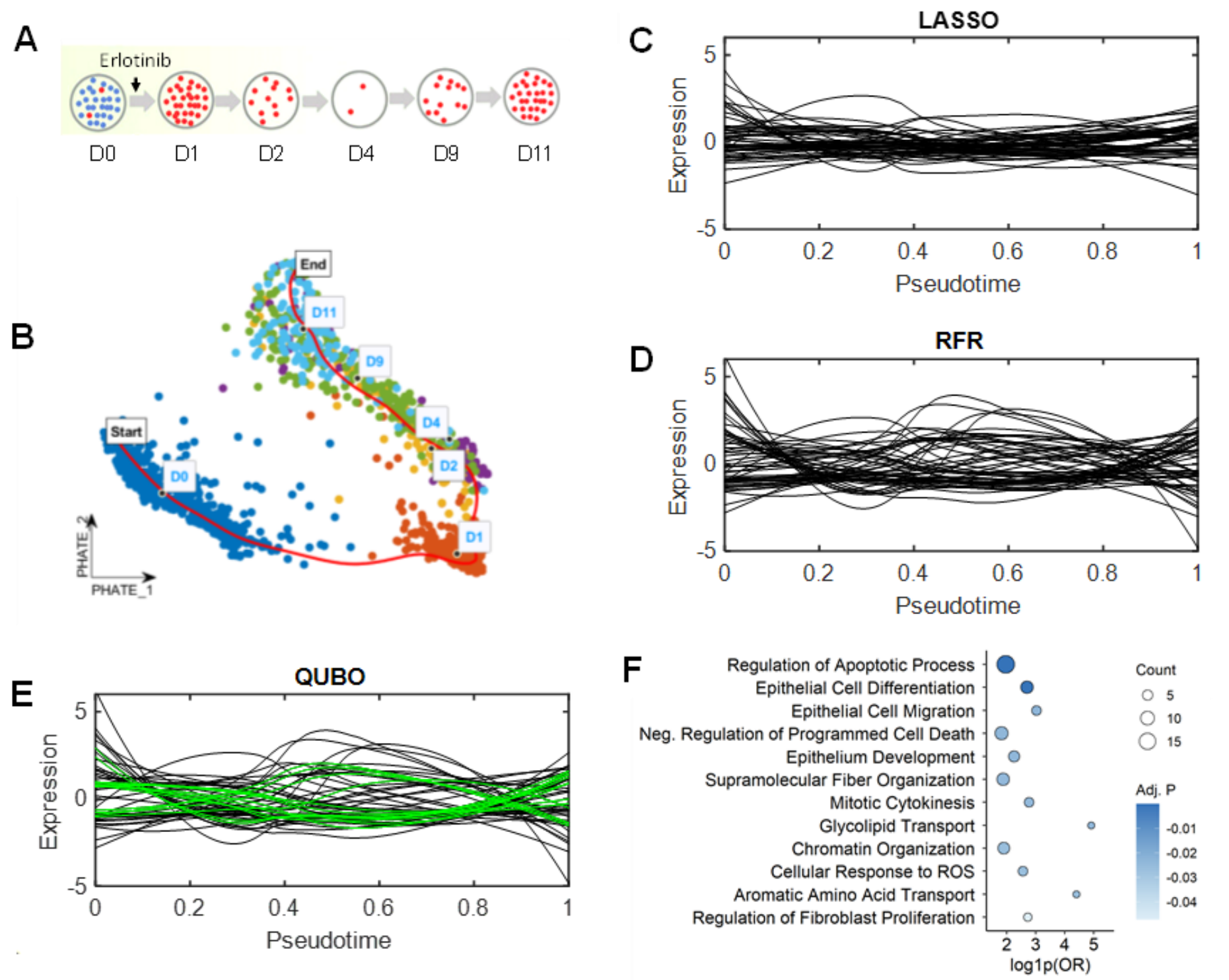

**Fig. 4. QUBO feature selection reveals genes associated with anticancer drug resistance.** **A.** Experimental schematic depicting PC9 cell samples treated with the anticancer drug erlotinib. D0 represents untreated cells, with D1 through D11 indicating the progressive duration of drug treatment. **B.** Two-dimensional visualization of cells using PHATE embedding, which enabled construction of the cell pseudotime trajectory. **C.** LOESS-smoothed expression profiles of the top 50 genes identified by LASSO regression, plotted against pseudotime. **D.** Expression profiles of top genes identified through RFR. **E.** Expression profiles of top genes identified by QUBO feature selection. Green lines indicate genes uniquely detected by the QUBO method that were overlooked by LASSO and RFR. **F.** Functional enrichment analysis of pathways associated with genes exclusively identified using the QUBO method.

## QUBO feature selection identifies key genes implicated in cancer drug resistance

Finally, we applied QUBO feature selection to another published scRNA-seq dataset from a study of drug tolerance in cancer (Aissa *et al*. 2021). This research focused on tyrosine kinase inhibitors (TKIs) as first-line anticancer drug, and the data was generated from the lung cancer cell line PC9 treated with a TKI drug (erlotinib) for 11 days (**Fig. 4A**). The scRNA-seq was done on cells at Day 0 (D0), Day 1 (D1), Day 2 (D2), Day 4 (D4), Day 9 (D9), and Day 11 (D11) of the treatments, and cells were pooled to form the pseudotime trajectory (**Fig. 4B**). LASSO, RFR, and QUBO methods were applied to identify feature genes and expression profiles of the top 50 genes each were plotted as a function of pseudotime, shown in **Fig. 4C,D,E**.

Among all top 50 genes, QUBO feature selection uniquely identified 28 genes overlooked by LASSO and RFR. These QUBO-exclusive genes demonstrated significant associations with drug response, cancer progression, and drug resistance mechanisms, revealing a critical set of genes implicated in complex cancer-related processes. Key findings include several well-established lung cancer marker genes: CCND1, BIRC5, and STMN1, primarily involved in cell cycle regulation and proliferation (Bao *et al.* ; Montalto and De Amicis ; Li *et al.*). ANLN and TPM3 are known to be implicated in cancer-related cytoskeletal organization and cell division (Armstrong *et al.* ; Naydenov *et al.*). GSTK1 (Aissa *et al.*) is known to be associated with erlotinib-specific metabolic reprogramming and oxidative stress response, and ALDH3A1 (Pu *et al.*) is a marker of paclitaxel resistance in lung cancer. To further elucidate the functional implications of the selected gene set, we performed pathway enrichment analysis (**Supplementary Table 2**). This analysis revealed significant enrichment of biological processes associated with drug response, cancer progression, and resistance (**Fig. 4F**). Notably, pathways involved in *negative regulation of cell migration* (Emran *et al.*) and *regulation of apoptotic processes* (Neophytou *et al.*) were significantly enriched, supporting their roles in modulating cancer cell survival and metastatic potential under drug treatment. The identification of the pathway *regulation of fibroblast proliferation* (Feng *et al.*) suggests a potential contribution to the tumor microenvironment's response to therapy. Moreover, the enrichment of *supramolecular fiber organization* and *chromatin organization* underscores the involvement of structural and epigenetic regulatory mechanisms in cancer (Sehgal and Chaturvedi). Finally, the pathways of *aromatic amino acid transport* and *epithelial cell differentiation* suggest that metabolic and differentiation state alterations contribute to drug response, which is a key feature of this time series lung epithelial cancer dataset developing drug resistance. These enriched pathways underscore the complex mechanisms of cancer cell survival, metastatic potential, and adaptive responses to therapeutic pressures.

The QUBO method demonstrated remarkable effectiveness in uncovering genes critical for mediating uncontrolled cell growth and survival under therapeutic stress. By identifying genes related to drug-specific resistance and metabolic adaptation, QUBO revealed insights into detoxification and redox balance mechanisms essential for cancer cell survival. Notably, the relationships between these selected genes and predictor variables could not be adequately explained by linear modeling. This highlights QUBO's unique capability to uncover complex, nonlinear interactions that traditional methods might miss, providing a more nuanced and interpretable understanding of genetic mechanisms in lung cancer progression and drug resistance.

## Discussion

In complex cellular systems, cells undergo differentiation, transitions, growth, division, and death while also responding to diverse environmental and intracellular signals that modulate gene expression programs. This complexity makes it challenging to study the internal controls that regulate specific cellular processes.

In this study, we demonstrate that QUBO feature selection, when applied to real-world scRNA-seq data, produces more insightful results compared to traditional feature selection methods. Our findings suggest that traditional machine learning methods, despite being cross-validated, do not necessarily yield high-quality feature selection sets. QUBO feature selection instead shows an incomparable robustness and stability, extracting genes with both linear and nonlinear expression profiles. Traditional methods often struggle with complex relationships between features, frequently overlooking valuable combinations. The quantum-mechanical nature of QUBO feature selection, however, enables exploration of intricate connections between features, identifying subsets that capture complex interactions and improve model performance. Furthermore, traditional feature selection becomes computationally prohibitive in high-dimensional datasets due to iterative processes. QUBO feature selection addresses this by translating the selection process into a form optimized for quantum hardware, allowing simultaneous exploration of numerous possibilities and effectively managing the computational complexity of high-dimensional feature selection.

Our study focuses on comparing QUBO with two traditional feature selection methods: LASSO and RFR. LASSO primarily assesses the importance of individual target variables, which can lead to the inclusion of redundant features containing similar information. In contrast, QUBO feature selection explores a broader solution space, allowing it to identify and eliminate redundant features while selecting the most informative ones, resulting in a more concise and efficient feature set. On the other hand, RFR, as a powerful ensemble learning method, captures complex nonlinear relationships using multiple decision trees. However, it suffers from black-box interpretability issues, high computational demands, and potential overfitting. In comparison, QUBO-based regression provides a more structured and interpretable approach by directly optimizing feature selection and coefficient values through combinatorial optimization.

While the choice of the $K$ parameter may be arbitrary and, if not selected carefully, could overlook important biological insights, our QUBO-based method consistently prioritizes the most impactful features regardless of the specific $K$ value. In bioinformatics, it is common to focus on the top 50 to 200 features, as seen in differential gene expression analysis. Similarly, we recommend selecting $K$ within this range when using our QUBO method to facilitate enrichment analysis and extract meaningful biological insights. Although a larger number of features can be extracted, focusing on the top-ranked ones often provides the most biologically relevant insights, as these features are more likely to be primary drivers of underlying biological processes.

By leveraging quantum or quantum-inspired solvers, QUBO-based regression offers potential exponential speedups for combinatorial optimization problems while naturally enforcing sparsity and regularization, making it particularly effective for high-dimensional regression challenges. Pseudotime is just an example of a target variable. Many other cellular continuous variables can be used in the QUBO feature selection for regression problems. This represents a significant step forward in the analysis and interpretation of scRNA-seq data along with other cell state measurements, offering a complementary approach to traditional statistical methods.

Beyond scRNA-seq data analysis, the implications of this study extend to broader areas of biological research. The successful application of quantum computing to complex biomedical data suggests vast potential for integrating quantum techniques into various domains. As quantum computing continues to evolve, its ability to address computational challenges in big data and high-dimensional datasets could drive groundbreaking advancements.

# Methods

## Processing of scRNA-seq data

For the case study of cell differentiation, the scRNA-seq data was obtained from a published study on a hESC-EC induction system (Xu *et al*. 2023). This system employed overexpression of the transcription factor FLI1 to induce hESC differentiation into ECs. This induction approach has been shown to be more efficient than cytokine stimulation (Xu *et al.* 2023). Data generated from the cell sample after 24-hour FLI1 induction was selected. The processed data contained 5,000 genes and 4,697 cells (consisting of about equal number of hESCs and ECs). For the case study of resistance development, the scRNA-seq data was obtained from a published study on anticancer drug resistance. The experiments were performed with non-small cell lung carcinoma PC9 cell lines. The time-course scRNA-seq was done before (i.e., day 0) and after cells were treated for 1, 2, 4, 9, or 11 days by adding erlotinib—the first-generation TKI drug. The scRNA-seq expression matrices were download from the GEO database using accession number GSE134839. The processed data contained 9,661 genes and 1,422 cells (643, 205, 133, 88, 217, and 136 for D0, D1, D2, D4, D9 and D11 samples, respectively). For both case studies, the scRNA-seq count matrices were processed in the similar manner. Briefly, matrices were imported into scGEAToolbox (Cai 2019) for quality control filtering and cell type or group examination. The default quality control filtering was applied with thresholds of library size of 1,000 minimum reads per cell, 15% maximum mitochondrial DNA ratio per cell, 15 minimum nonzero cells per gene, and 500 minimum nonzero genes per cell. PHATE (Moon *et al.* 2019) was used to embed and visualize cells. The splinefit algorithm (Cai 2019) was used for trajectory inference to estimate the pseudotime of cells. The estimated pseudotime was used as the target variable $\boldsymbol{T}$ for constructing the cross-entropy terms in the mutual information calculation in Equations (5) and (6). For regression analysis, the gene-by-cell count matrix was transformed using Pearson residual transformation (Lause *et al.* 2021).

The QUBO formulation, detailed in the next section, was then applied to identify $k = 50$ informative features (genes) from the processed data. The computation was conducted using quantum annealing via the Ocean SDK hybrid solver (Stogiannos *et al.* 2022) in D-Wave's quantum annealer, as well as using the iterated tabu search algorithm (Palubeckis 2006) in the MATLAB quantum computing package.

## Transition from Ising model to QUBO problem

The Ising model, derived from statistical mechanics and inspired by ferromagnetism, employs a Hamiltonian operator to define an energetic state or cost function as follows,

$$\hat{H}(\boldsymbol{\sigma}) = -\sum_{i=1}^{N} h_i \sigma_i - \sum_{i=1}^{N} \sum_{i<j}^{N} J_{ij} \sigma_i \sigma_j . \quad (1)$$

Here $\boldsymbol{\sigma}$ represents the spin state operator, indicating the "up" or "down" magnetic states (Brooke *et al.* 1999). This can naturally be translated to classical binary information, making it suitable for translating classical information to quantum information. The Ising model's Hamiltonian in Equation (1) defines an energy state, analogous to a cost function in optimization problems, and uses the spin state operator and spin-spin coupling constants $J_{ij}$ to capture interaction between spins. An external magnetic field $h_i$ tunes the ground states for a particular problem. The expected value of the Hamiltonian determines the energy or cost function ($\langle H \rangle = E$). The parametric dependence on $J_{ij}$ and $h_i$ defines the energy/cost function value, while the expectation value of the Hamiltonian carries implicitly a probabilistic quantum behavior. Interestingly, the Ising model's mathematical formulation resembles that of QUBO problem, which is a combinatorial optimization problem commonly encountered in various fields (MÜCKE *et al.* 2023). QUBO feature selection has become noticeable, since there are some applications for the quantum computing area such as ranking and classifying QA-FS (DACREMA *et al.* 2022) and feature selection applied to recommender system (NEMBRINI *et al.* 2021).

The QUBO problem aims to minimize a cost function $f(\boldsymbol{x}) = \boldsymbol{x}^T \boldsymbol{Q} \boldsymbol{x}$, where $\boldsymbol{Q}$ is an upper triangular matrix, and $\boldsymbol{x} = (x_1, \ldots, x_N)^T$ contains $x_i$ binary elements. The QUBO matrix $\boldsymbol{Q}$ encodes the problem's constraints and objectives, while the column vector $\boldsymbol{x}$ represents the variables to be optimized. Usually, $f(\boldsymbol{x})$ diagonal elements $(q_i Q_{ii} q_i)$ can be simplified due to binary representation $(q_i q_i = q_i)$ as follows,

$$f(\boldsymbol{x}) = \sum_{i=1}^{N} Q_{ii} x_i + \sum_{i=1}^{N} \sum_{i<j}^{N} Q_{ij} x_i x_j. \tag{2}$$

The annealing process minimizes the energetic configuration, akin to finding the ground spin state for the encoded problem. As previously noted, quantum annealing can perform cost function minimization more effectively than classical computing. Interestingly, this objective minimization is analogous to the least squares problem, where matrices $\boldsymbol{A}$, $\boldsymbol{x}$, and $\boldsymbol{b}$ are used to minimize $\|\boldsymbol{A}\boldsymbol{x} - \boldsymbol{b}\|^2$ for an optimal $\boldsymbol{x}$. Although $\|\mathbf{Ax} - \mathbf{b}\|^2$ and the QUBO matrix $\boldsymbol{Q}$ are described using linear equations, they naturally represent quadratic forms (JUN 2024).

## QUBO feature selection model construction

The feature selection problem aims to identify a subset $S \subset [n]$ of informative features from the original set $[n] = \{1,2, \ldots, n\}$, where $n$ is the total number of features. Suppose a dataset $\boldsymbol{D} \coloneqq \{(\boldsymbol{x}^i, \boldsymbol{\tau}^i)\}_{i \in [n]}$ containing $n$ features and $p$ observations, where $\boldsymbol{x}^i \in \boldsymbol{X} \subseteq \mathbb{R}^p$ represents the feature vectors and $\boldsymbol{\tau}^i \in \boldsymbol{T} \subseteq \mathbb{R}^p$ represents the target variable. The goal is to preserve the nature of the data while reducing its size for better interpretation and reduced complexity by obtaining a subset of the original dataset $D_S = \{(\boldsymbol{x}_S^i, \boldsymbol{\tau}^i)\}_{i \in [n]}$ with $\boldsymbol{x}_S^i \in \boldsymbol{X}_S \subseteq \mathbb{R}^p$. In our implementation, we aim to accurately describe the underlying biological processes with a representative set of features/genes.

Interestingly, this task can be formulated as an optimization problem with a cost function. We propose a novel QUBO-based feature selection approach (MÜCKE *et al.* 2023) for scRNA-seq data. The QUBO formulation seeks the optimal feature vector $\boldsymbol{F}$ that minimizes the cost function $\boldsymbol{Q}$ as follows,

$$\boldsymbol{F}^* \coloneqq arg_{\boldsymbol{F} \in \{0,1\}^n} \, min \, \boldsymbol{Q}(\boldsymbol{F}, \alpha; \boldsymbol{I}, \boldsymbol{R}), \tag{3}$$

where $\boldsymbol{F}^*$ is a binary vector containing $F_i^* = 1$ for selected features. Here, $\boldsymbol{I}$ and $\boldsymbol{R}$ are parameters defining the cost function. $\boldsymbol{I}$ represents the MI within feature $\boldsymbol{x}^i$ and target $\boldsymbol{\tau}^i$, such that $I_i \in \boldsymbol{I} \subseteq \mathbb{R}^n$. $\boldsymbol{R}$ represents the MI within features $\boldsymbol{x}^i$ and $\boldsymbol{x}^j$, such that $R_{ij} \in \boldsymbol{R} \subseteq \mathbb{R}^{n \times n}$. $\alpha$ is a parameter between 0 and 1 used to balance the solution. Our QUBO cost function $\boldsymbol{Q}$ is defined as follows,

$$Q(F, \alpha; I, R) \coloneqq -\alpha \sum_{i=1}^{n} I_i F_i + (1-\alpha) \sum_{i,j=1}^{n} R_{ij} F_i F_j. \tag{4}$$

The $I_i = I(\boldsymbol{x}^i; \boldsymbol{\tau}^i)$ [Equation (5)] is the $i$-th importance element. The redundancy element $R_{ij} = I(\boldsymbol{x}^i; \boldsymbol{x}^j)$ [Equation (6)] is the MI between the $i$-th and $j$-th features. Since a self-feature is never redundant, we set $R_{ii} = 0$. MI values are positive, indicating strong interactions with higher values and no interaction near zero. This framework evaluates the importance of the target $\boldsymbol{\tau}^i$ in relation to individual features, where crosstalk between features reduces the importance of some variables. This determines each feature's overall relevance to the response variable $\boldsymbol{\tau}^i$, is balanced by the parameter $\alpha$ $(0 \le \alpha \le 1)$.

One challenge is that directly estimating mutual information from real-world data is difficult due to the requirement for the joint probability mass function of the features and target variables. To address this challenge, we employed a binned discretization approach. Here, each feature was divided into $B$ bins using quantiles, ensuring a fair distribution of the data across the bins. Similarly, we applied discretization to our target variable since it is a continuous variable. The binning process is as follows:

- Quantile calculation: For each feature $\boldsymbol{x}^i$, we calculate the $(B+1)$ quantiles $q_i^L$ for $L \in \{0,1,\dots,B\}$. These quantiles create the boundaries of the bins.
- Binning: Each bin $\beta_i^L$ is defined as the interval $[q_i^{L-1}, q_i^L)$ for $L \in \{1,2,\dots,B-1\}$, and the final bin $\beta_i^B$is $[q_i^{B-1}, q_i^B]$.
- Assigning bins: Each feature value $x_j^i$ is assigned to a bin based on which interval it falls into. For instance, if $x_j^i$ falls between $q_i^{L-1}$ and $q_i^L$, it is assigned to bin $L$ $(x_j^i \in \beta_i^L)$.

The discretized features $\widehat{\boldsymbol{x}}^i$ and target variable $\widehat{\boldsymbol{\tau}}^i$ are integrated into the discretized data $\widehat{D} = \{(\widehat{\boldsymbol{x}}^i, \widehat{\boldsymbol{\tau}}^i)\}_{i\in[n]}$, where $\widehat{\boldsymbol{x}}^i \in [\beta_{i,x}] \subseteq \mathbb{R}^B$ and $\widehat{\boldsymbol{\tau}}^i \in [\beta_{i,\tau}] \subseteq \mathbb{R}^B$ with $\beta_{i,\boldsymbol{k}} = \{\beta_{i,\boldsymbol{k}}^1, \dots, \beta_{i,\boldsymbol{k}}^B\}$ for all $i \in [n]$ . This binning approach allows us to approximate the information entropy across features and target variables. Thus, importance ($\boldsymbol{I}$) and redundancy ($\boldsymbol{R}$) are described as follows,

$$I_i \coloneqq I(\boldsymbol{x}^i; \boldsymbol{\tau}^i) \approx \sum_{\widehat{\boldsymbol{x}}^i \in [\beta_{i,x}]} \sum_{\widehat{\boldsymbol{\tau}}^i \in [\beta_{i,\tau}]} \hat{p}(\widehat{\boldsymbol{x}}^i, \widehat{\boldsymbol{\tau}}^i) \, log\left(\frac{\hat{p}(\widehat{\boldsymbol{x}}^i, \widehat{\boldsymbol{\tau}}^i)}{\hat{p}(\widehat{\boldsymbol{x}}^i)\hat{p}(\widehat{\boldsymbol{\tau}}^i)}\right), \tag{5}$$

$$R_{ij} \coloneqq I(\boldsymbol{x}^i; \boldsymbol{x}^j) \approx \sum_{\widehat{\boldsymbol{x}}^i \in [\beta_{i,x}]} \sum_{\widehat{\boldsymbol{x}}^j \in [\beta_{j,x}]} \hat{p}(\widehat{\boldsymbol{x}}^i, \widehat{\boldsymbol{x}}^j) \, log\left(\frac{\hat{p}(\widehat{\boldsymbol{x}}^i, \widehat{\boldsymbol{x}}^j)}{\hat{p}(\widehat{\boldsymbol{x}}^i)\hat{p}(\widehat{\boldsymbol{x}}^j)}\right). \tag{6}$$

Equations (5) and (6) involve summation over discretized bins and calculate the log-ratio between joint and marginal probabilities estimated from discretized data $\widehat{D}$. The discretized empirical probabilities mass function is defined as,

$$\hat{p}(\widehat{\boldsymbol{x}}^i, \widehat{\boldsymbol{\tau}}^i) \coloneqq \frac{1}{n} \sum_{(\widehat{\boldsymbol{x}}', \widehat{\boldsymbol{\tau}}') \in \widehat{D}} \mathbb{I}_{\{\widehat{\boldsymbol{x}}^i = \widehat{\boldsymbol{x}}' \wedge \widehat{\boldsymbol{\tau}}^i = \widehat{\boldsymbol{\tau}}'\}}, \tag{7}$$

with an indicator function,

$$\mathbb{I}_{\{P\}} \coloneqq \begin{cases} 1 \text{ if P is true} \\ 0 \text{ otherwise} \end{cases}, \quad (8)$$

defined for logical statements $P$. A semi-last remark, Equation (4) can be simplified as $Q_{ij}(\alpha) = R_{ij} - \alpha(R_{ij} - \delta_{ij} I_i)$, where the Kronecker delta $\delta_{ij}$ is unity if $i = j$ and 0 otherwise. Interestingly, this QUBO matrix $\boldsymbol{Q}$ can find an optimal or a quasi-optimal solution to the feature selection problem. It also avoids gauge problems commonly encountered in QUBO formulations. In our simulations, we consider one predictor for all features $\boldsymbol{\tau}^i \equiv \boldsymbol{T}$. Importantly, while the QUBO model is unconstrained, additional penalty terms could be introduced to customize the energy landscape for specific requirements ($Q' = Q^{problem} + MQ^{constraint}$).

## Code Availability

Code implementing the method described in this paper is available at https://github.com/cailab-tamu/QUBO_feature_selection along with example data.

## Acknowledgment

We are grateful to Dr. Liang Hu for sharing the hESC-EC scRNA-seq data. We acknowledge the use of advanced computing resources provided by Texas A&M High Performance Research Computing in conducting parts of this research.

## Funding

This study was funded by the U.S. Department of Defense (DoD, GW200026) and the National Institute for Environmental Health Sciences (P30 ES029067) for J.J.C, Allen Endowed Chair in Nutrition & Chronic Disease Prevention for R.S.C., and the Cancer Prevention & Research Institute of Texas (CPRIT, RP230204) for J.J.C. and R.S.C.